\documentstyle[11pt]{article}
\input{epsf}

\def\spacing #1{\small \renewcommand{\baselinestretch}{#1} \normalsize}

\newcounter{nr}
\newcounter{subnr}[nr]
\newcounter{subsubnr}[subnr]

\def\thebibliography#1{\par \vspace{2ex}
 \begin{center} {REFERENCES} \end{center}
 \nopagebreak \vspace*{-1.3ex} \list
 {\arabic{enumi}.}{\settowidth\labelwidth{3ex}\leftmargin\labelwidth
 \advance\leftmargin\labelsep
 \usecounter{enumi}}
 \def\newblock{\hskip .11em plus .33em minus -.07em}
 \sloppy
 \sfcode`\.=1000\relax}

\begin{document}
\spacing{0.9091}
\parskip=0ex \parindent=0ex
\small

\twocolumn[{%
\begin{center}
    {\large \bf  SEISMIC INVESTIGATION OF THE SOLAR STRUCTURE
USING GONG FREQUENCIES}\\[4EX]   
    {\normalsize \bf
S.C. Tripathy$^{1}$, H.M. Antia$^2$, F. Hill$^3$ and A. Ambastha$^1$ 	
}\\[4ex]
    \begin{minipage}[t]{16cm}
      \begin{tabbing}               
$^1$ Udaipur Solar Observatory, Physical Research Laboratory, PO
Box No. 198, Udaipur 313 001, India.\\
$^2$ Tata Institute of Fundamental Research, Mumbai 400005, India.\\
$^3$ National Solar Observatory, 950 Cherry Avenue, Tucson, USA.\\
      \end{tabbing}
    \end{minipage}
\end{center}}]



We investigate the effect of equation of state and
convection formalism on the sound speed in solar interior by
using the asymptotic inversion technique of Christensen-Dalsgaard, Gough \&
Thompson (1989). In this technique the sound speed difference between
two solar models is estimated from the known frequency differences
using,
\begin{equation}
S(w){\delta\omega \over\omega} = S(w) {\omega_0 - \omega \over \omega_0} = H_1 (w) + H_2 (\omega),
\end{equation}
where
\begin{equation}
S(w) = \int_{r_t}^{R_\odot} \left( 1 - {c_0^2\over w^2r^2}\right) {dr\over c_0},
\end{equation}
$\omega_0$ is the frequency of p-mode in the reference model,
$\omega$ is the observed frequency for the same mode and $w =
\omega/(\ell+1/2)$. The function $H_1(w)$ which is related to
the sound speed difference between the model and the Sun,
provides another measure of the sound speed difference.
Similarly, $H_2(\omega)$ which reflects the differences
in the surface layers is better suited to test different
formulations of solar convection theory.
The sound speed difference between the Sun and model can be found using the relation
\begin{equation}
{\delta c\over c} = {c_0 - c \over c} = - {2r\over \pi} {da\over dr} \int_{a}^{a_s} {{dH_1\over dw} w dw  \over (a^2-w^2)^{1/2}},
\end{equation}
where $a=c/r$ and  $a_s = a(R_\odot)$.
As an independent diagnostic for differences in the surface layers
we also consider the scaled frequency difference between the models and the Sun
defined by $\Delta_{n,l}$ = $Q_{n,l} \delta\omega$, where
$Q_{n,l} = E_{n,l}/\bar{E}_0 (w_{n,l})$, $E_{n,l}$ is the mode energy and
$\bar{E}_0 (\omega_{n,l})$ is the value of $E_{n,l}$ for $\ell$ = 0
interpolated to the frequency $\omega_{n,l}$.

\begin{table}[htbp]
  \caption[]{\label{tab:xx}     
Properties of solar models}     			
  \begin{center}
\footnotesize
\begin{tabular}{cccccc}
    \hline
Model&Conv&EOS&$X$&$r_d/R_\odot$&{$\rho_{\rm c}$}\\ 
\hline
\hline
M1&CM&OPAL&0.7275&0.71643&153.0\\ 
M2&MLT&OPAL&0.7275&0.71659&153.1\\ 
M3&CM&MHD&0.7267&0.71761&153.2\\ 
M4&MLT&MHD&0.7267&0.71776&153.2\\ 
\hline
   \end{tabular}
  \end{center}
\end{table}

For this investigation we construct different solar models
having identical physics except for the equation of state and treatment of
convection.
All the models use OPAL opacities (Rogers \& Iglesias 1992)
and incorporate diffusion of helium and heavy elements
using the hydrogen abundance profile from Bahcall and Pinsonneault
(1992) and heavy element abundance profile from Proffitt (1994).
The convective flux has been calculated using either the usual Mixing
Length Theory (MLT) or  Canuto \& Mazzitelli (1991) (CM) formulation.
The properties of these models with heavy element abundance $Z$=0.019,
are summarised in Table 1.
These models use either MHD (Mihalas et al.~1988) or OPAL (Rogers et al.~1996)
equations of state.

The  error estimate due to the uncertainties in the observed
frequencies is obtained by  simulating
30 sets of frequencies where random errors
with standard deviation quoted by the observers
are added to the model frequencies.
The variance in the computed values of relative
sound speed difference gives the error estimate.
The frequencies
used are from the GONG network (Hill et al.~1996) spanning month 7, month 8
and those derived from the averaged spectra of months 4 to 7.

\begin{figure}[htbp]
\epsfxsize=7cm
\centerline{\epsfbox[29 19 200 180]{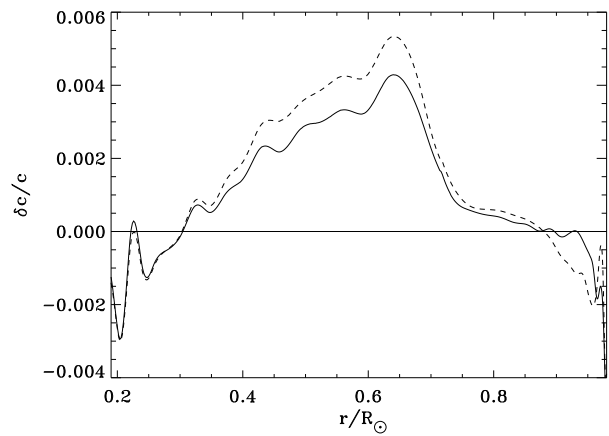}}
    \caption[]{\label{fig:myfig1}	
Relative sound speed difference between the Sun and models
constructed using OPAL (solid line) and MHD (dashed line) equations
of state. The hump near $r/R_{\odot}$ = 0.95 is visible for the MHD
equation of state}			
\end{figure}

From the tests performed on inversion results it is clear that the
differential asymptotic inversion technique is capable of providing
reasonable results in the range $0.2R_\odot<r<0.95R_\odot$.
Figure~1 shows the relative difference in sound
speed between the Sun and models M1 and M3.
It is evident from the figure that the sound speed in model M1 using
OPAL equation of state is closer to that in the Sun as compared to model M3
which uses MHD equation of state.
Most of this difference in the radiative interior
has come from the difference in depth of the convection
zone between the two models. We can also notice a hump around $r=0.95R_\odot$
inside the convection zone in MHD model.
Thus it appears that the OPAL equation of state better describes
the solar material  which also confirms earlier results
(Basu \& Antia 1995a; D\"appen 1996).

The two formalisms for stellar convection were tested by
performing inversions using models M1 and M2. It appears that the model M1,
which uses
the CM formulation for calculating the convective flux is closer to
the Sun as compared to the MLT model. However, a definite
conclusion can not be made on the basis of inversion results as
most of the difference occurs in the surface layers where the
inversions are not particularly reliable.

\begin{figure}[htbp]
\epsfxsize=7cm
\centerline{\epsfbox[29 19 200 170]{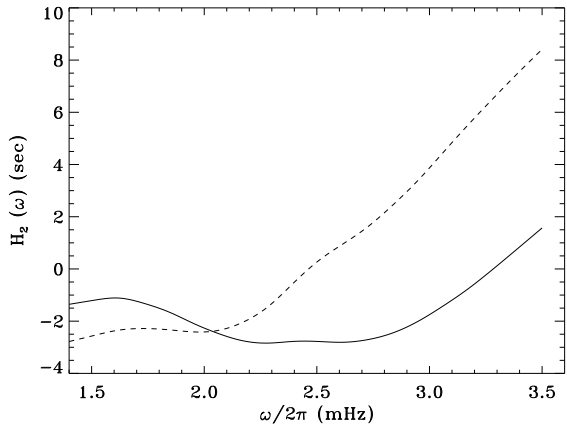}}
    \caption[]{\label{fig:myfig2}	
The function $H_2(\omega)$ between the Sun and
models constructed with different treatment of convection.
The sold line represents CM model and the dashed line MLT model.}
\end{figure}

Figure~2 shows $H_2(\omega)$ as a function of $\omega$/2$\pi$ for models
M1 and M2.
We find that model M1 using CM formalism
has a significantly smaller variation 
than model M2 using MLT formulation. In
particular, in CM model $H_2(\omega)$  is practically flat for
$\omega < 3$ mHz probably indicating that CM models are  closer to
the Sun than models constructed using MLT.
The scaled frequency difference between the models M1 and
M2 and the Sun is plotted in Figure~3.  It is evident that the
scaled frequency difference between the CM model and the Sun is smaller
than that between the MLT model and the Sun. Therefore, we conclude that
the models
using CM formulation for convection are closer to the Sun and
further confirm earlier results (Paterno {\it et al.} 1993, Basu \& Antia 1994, 1995b).

\begin{figure}[htbp]
\epsfxsize=7cm
\centerline{\epsfbox[29 19 200 170]{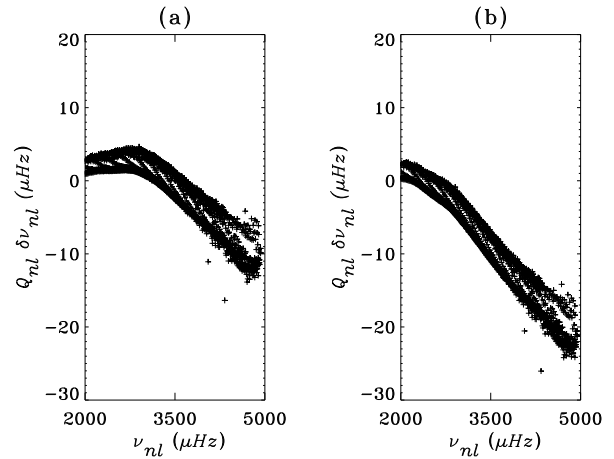}}
    \caption[]{\label{fig:myfig3}	
The scaled frequency difference between the Sun and (a) model
M1, and (b) model M2.} 
\end{figure}

\medskip
Acknowledgments:
This work utilizes data obtained by the Global Oscillation
Network Group (GONG) project, managed by the National Solar Observatory, a
Division of the National Optical Astronomy Observatories, which is operated by
AURA, Inc. under a cooperative agreement with the National Science
Foundation. One of the authors (SCT) acknowledges financial support from DST, Govt. of India.

\end{document}